\documentclass[a4paper,12pt]{article}
\usepackage{epsfig}
\newcommand{\Str}{\ensuremath{\mathop{\rm Str}\nolimits}}
\newcommand{\tr}{\ensuremath{\mathop{\rm tr}\nolimits}}
\newcommand{\ath}{\ensuremath{\mathop{\mathrm{arctanh}}}}
\newcommand{\mn}{{\mu\nu}}

\newcommand{\cE}{{\cal E}}

\newcommand{\cA}{{\cal A}}

\newcommand{\cR}{{\cal R}}
\newcommand{\duF}{\tilde{F}}
\newcommand{\dif}[1]{\frac{\partial}{\partial #1}}

\newcommand{\be}{\begin{equation}}
\newcommand{\ee}{\end{equation}}
\newcommand{\bea}{\begin{eqnarray}}
\newcommand{\eea}{\end{eqnarray}}

\begin{document}
\title{
  \begin{flushright} \begin{small}
  DTP--MSU/00-08 \\ hep-th/0006242
  \end{small} \end{flushright}
\vspace{2.cm}
%%%%  Title  %%%%
{\bf Sphaleron glueballs in NBI theory with symmetrized trace }%
\bigskip}
%%%%%  Authors  %%%%
\author{ V.V. Dyadichev\thanks{Email: rkf@mail.ru}
      and  D.V. Gal'tsov\thanks{Email: galtsov@grg.phys.msu.su} \\ \\
{\it Department of Theoretical Physics,}   \\
       {\it  Moscow State University, 119899, Moscow, Russia}
}
%%%%%  Date  %%%%
\date{\today}
\maketitle

\begin{abstract}
We derive a closed expression for the $SU(2)$ Born-Infeld action with
the symmetrized trace for static spherically symmetric purely magnetic
configurations. The lagrangian is obtained in terms of elementary functions.
Using it, we investigate glueball solutions to the flat space NBI theory
and their self-gravitating counterparts. Such solutions, found previously
in the NBI model with the 'square root -- ordinary trace' lagrangian, are
shown to persist in the theory with the symmetrized trace lagrangian as well.
Although the symmetrized trace NBI equations differ substantially from those
of the theory with the ordinary trace, a qualitative picture of
glueballs remains essentially the same. Gravity further reduces the
difference between solutions in these two models, and, for sufficiently large
values of the effective gravitational coupling, solutions tends to the same
limiting form. The black holes in the NBI theory with the symmetrized trace
are also discussed.
\end{abstract}
\newpage
%%%%%%%%%%%%%%%%%%%%%%%%%%%%%%%%%%%%%%%%%%%%%%%%%%%%%%
\section{Introduction}
%%%%%%%%%%%%%%%%%%%%%%%%%%%%%%%%%%%%%%%%%%%%%%%%%%%%%%
Recent development in the superstring theory suggests that the
low-energy dynamics of $N$ coincident $D$-branes is described by
the $SU(N)$ Yang-Mills theory governed by the Born-Infeld type
action \cite{GiKu98}. A precise definition of such non-Abelian
Born-Infeld (NBI) action was the subject of a vivid discussion
during past few years \cite{Ts97,GaGoTo98,Br98,BrPe98,Pa99,Za00},
for an early discussion see \cite{Ha81}. An
ambiguity is encoded in specifying the trace operation over the
gauge group generators. Formally a number of possibilities can be
envisaged. Starting with the determinant form of the $U(1)$
Dirac-Born-Infeld action
\be  \label{det}
S=\frac{1}{4\pi}\int
\left\{1-\sqrt{-\det(g_{\mu\nu}+F_{\mu\nu})}\right\}d^4 x
\ee
one can use the usual trace, the symmetrized  or
antisymmetrized \cite{Ts97} trace, a calculation of the determinant both
with respect to Lorentz and the gauge matrix indices \cite{Pa99}.
Alternatively one can start with the 'square root' form, which is
most easily derived from (\ref{det}) using the identities
\bea
&\det(g_{\mu\nu}+F_{\mu\nu})=\det(g_{\mu\nu}-F_{\mu\nu})=
\det(g_{\mu\nu}+i{\tilde F}_{\mu\nu})=&\\
&\det(g_{\mu\nu}-i{\tilde F}_{\mu\nu})=
\left[\det((g_{\mu\nu}-F^2_{\mu\nu})(g_{\mu\nu}+{\tilde
F}^2_{\mu\nu}))\right] ^{1/4}&,
\eea
where $F^2_{\mn}=F_{\mu\alpha}{F^{\alpha}}_\nu$ (similarly for $\duF_\mn$),
and
\bea \label{id}
F_{\mu\alpha}{F^\alpha}_\nu-\duF_{\mu\alpha}\duF^\alpha{}_\nu&=&
\frac12 g_{\mu\nu} F_{\alpha\beta}F^{\alpha\beta},\nonumber\\
F_{\mu\alpha}\duF^\alpha{}_\nu&=& -\frac14
g_{\mu\nu}F_{\alpha\beta}\duF^{\alpha\beta},
\eea
leading to the equality
\be
\sqrt{-\det(g_{\mu\nu}+F_{\mu\nu})}=
\sqrt{-\det(g)}\;\sqrt{1+\frac12 F^2 -\frac{1}{16}(F\tilde F)^2},
\ee
with $F^2=F_\mn F^\mn,\; F\duF=F_\mn\duF^\mn$. For a non-Abelian
gauge group the relations (\ref{id}) are no longer valid, so there
is no direct relationship between the 'determinant' and the
'square root' form of the lagrangian. Therefore the latter can  be
chosen as an independent starting point for a non-Abelian
generalization. There is, however, a particular trace operation --
symmetrized trace -- under which generators commute with each
other and therefore both forms of the lagrangian remain
equivalent. This definition is favored by the adiabaticity
argument, as was clarified by Tseytlin \cite{Ts97}. Restricting
the validity of the effective action  by the no-derivative
approximation, in the non-Abelian case one has to drop the
commutators of the matrix-valued $F_{\mu\nu}$ since they can be
reexpressed through the derivatives of  $F_{\mu\nu}$. This
corresponds to the following definition of the action
\be  \label{Strdet}
S=\frac{1}{4\pi}\Str\int
\left\{1-\sqrt{-\det(g_{\mu\nu}+F_{\mu\nu})}\right\}d^4 x
\ee
where symmetrization applies to the field strength (not
to potentials \cite{Ts97}). This action reproduces an exact
string theory result for non-Abelian gauge fields up to
$\alpha'^2$ order. Although there is no reason to believe that
this will be true in higher orders in $\alpha'$, the $\Str$ action is
an interesting model providing a minimal generalization of the
Abelian action \cite{Ts97}.

Some general technique was developed \cite{Za00} to deal with the
symmetrized products of gauge generators in a symbolic form, but
for many purposes it is more desirable to have the lagrangian
explicitly. In the general case an evaluation of the action can be
performed through an expansion of the square root in powers of
$F_{\mu\nu}$, then the symmetrized trace over generators can be
computed explicitly. The next step, resummation of the series into
a closed expression, is very problematic. However this can be done
if we restrict the field by some particular configurations (the
ansatz has to be consistent with the full equations of motion).
One example of this kind is the computation of an explicit action
for $D0$-branes in three dimensions \cite{ArFeKo98}. Here we
present  such a calculation in the four-dimensional $SU(2)$ theory
restricting the field configurations by requirements of the
spherical symmetry and staticity. Such configurations are
encountered in the study of magnetic monopoles and other solitons
in the NBI theory (we use a term 'soliton' in a wide sense
including unstable sphaleron solutions).

Soliton solutions to non-Abelian Born-Infeld theory were discussed
recently in a number of papers  using both the square root action
with the ordinary and the symmetrized trace
\cite{GrMoSc99,Gr99,Pa99,Tr99,LaTo99}, in the latter case,
however, only perturbatively. It was shown that in the Born-Infeld
theory, apart from monopoles and instantons, which also exist in
the theory with the usual Yang-Mills quadratic action, new
particle-like solutions  of the glueball type are brought to
existence \cite{GaKe99}. For these solutions the full non-linear
structure of the NBI lagrangian is essential, so earlier they
could be explored only within the model with the ordinary trace.
Here we show that these solutions persist in the theory with the
action (\ref{Strdet}). Although the explicit $\Str$ lagrangian
looks very differently from that with the ordinary $\tr$, the
glueball solutions remain qualiatively the same. We also study the
static spherically symmetric NBI system coupled to gravity and
demonstrate that gravity further reduces the difference between
solutions obtained in these two models.

%%%%%%%%%%%%%%%%%%%%%%%%%%%%%%%%%%%%%%%%%%%%%%%%%%%%%%%%%%%%%%%%
\section{Symmetrized trace NBI action for static $SO(3)$-symmetric fields}
%%%%%%%%%%%%%%%%%%%%%%%%%%%%%%%%%%%%%%%%%%%%%%%%%%%%%%%%%%%%%%%%
We define the NBI Lagrangian in four spacetime dimensions as
\be \label{LNBI}
L_{NBI}=\frac{\beta^2}{4\pi} \Str\left(1- \sqrt{-\det\Bigl(g_{\mu\nu}+
\frac{1}{\beta}F_{\mu\nu}\Bigr)}\right)=
\frac{\beta^2}{4\pi}\Str (1-\cR),
\ee
where
\be
\cR=\sqrt{1+\frac{1}{2\beta^2}F_{\mu\nu}F^{\mu\nu}
-\frac{1}{16\beta^4}(F_{\mu\nu}{\tilde F}_{\mu\nu})^2},
\ee
with the parameter $\beta$ of the dimension of length$^{-2}$ (the
'critical field'). The normalization of the gauge group generators
is chosen as follows
\be
F_\mn=F^a_\mn t_a,\quad \tr t_a t_b =\delta_{ab}.
\ee

The symmetrized trace of the product of $p$ matrices is defined as
\be
\Str(t_{a_1}\dots t_{a_p})\equiv\frac{1}{p!} \Str\left(t_{a_1}\dots
t_{a_p} + \mbox{all permutations}\right),
\ee
and it is understood that the general matrix function like
(\ref{LNBI}) has to be series expanded. It has to be noted that
while under the $\Str$ operation the generators  obviously can be
treated as commuting objects, the gauge algebra can not be
applied, i.e. $\tau_a^2\neq 1$ until the symmetrization of the
expansion is completed.

A general  $SO(3)$ symmetric $SU(2)$ gauge field  is
described by the Witten's ansatz
\be\label{wittans}
\sqrt{2}A = a_{0}t_{1}\; dt + a_{1}t_{1}\; dr +\{ \tilde w\; t_{2} -
(1-w)\; t_{3}\}\ d\theta +\{(1-w)\ t_{2} + \tilde w \;
t_{3}\}\sin\theta\; d\phi,
\ee
where the functions $a_0,\,a_1,\, w,\, \tilde w$ depend on $r,t$
and $\sqrt{2}$ is introduced to maintain the standard
normalization. Here we use a rotated basis $t_i,\, i=1,2,3$ for
the $SU(2)$ generators defined as
\be
t_1=n^{a}\tau^{a}/\sqrt{2},\quad
t_{2}=\partial_{\theta}t_{1},\quad \sin\theta
t_{3}=\partial_{\varphi}t_{1},
\ee
where $n^{a}=(\sin\theta  \cos\varphi,\sin\theta\sin\varphi,
\cos\theta)$, and $\tau^a$ are the Pauli matrices. These
generators obey the commutation relations
$[t_i,t_j]=\frac{1}{\sqrt{2}}\epsilon_{ijk}t_k$.

From four functions entering this ansatz one can be gauged away.
In the static case we can further reduce the number of independent functions
to two, while static purely magnetic configurations can be described
by a single variable  $w(r)$. Here we deal with this simplest case
\be\label{Acompts}
\sqrt{2}  A_{\theta}=-(1-w)t_3,\qquad
\sqrt{2}A_{\varphi}=\sin\theta(1-w)t_2,\qquad A_r=A_t=0.
\ee
The field strength tensor has the following non-zero components
\be \label{Fcompts}
 \sqrt{2} F_{r \theta}=w't_3,\qquad  \sqrt{2}F_{r \varphi}=-\sin\theta w' t_2,
  \qquad \sqrt{2}F_{\theta \varphi}=\sin\theta(w^2-1)t_1,
\ee
where prime denotes derivatives with respect to $r$.
 For purely magnetic configurations the second term under
the square root is zero, while the substitution of (\ref{Fcompts})
gives
\be
\cR^2=1+\frac{(1-w^2)^2}{\beta^2 r^4} t_1^2 +\frac{{w'}^2}{\beta^2
r^2} (t_2^2+t_3^2).
\ee
Now we have to expand the square root in a triple series in terms
of the even powers of generators $t_1, t_2, t_3$. This can be
achieved applying the formula
\be \label{sqrtexp}
  \sqrt{1+x}=1-2\sum_{m=1}^{\infty}\frac{(2m-2)!}{m!(m-1)!}
  {\left(-\frac{x}{4}\right)}^{m},
\end{equation}
and further repeatedly using binomial expansions. Finally we
obtain
\be
\label{Row}
L_{NBI}= \frac{\beta^2}{4\pi} \sum_{i+j+k\geq
1}^{\infty}\frac{(-1)^{i+j+k}(2i+2j+2k-2)!}{(i+j+k-1)!i!j!k!}\;
  \left(\frac{V}{2}\right)^{2i}\left(\frac{K}{2}\right)^{2j+2k}\;f(i,j,k),
\ee
where the sum over all positive $i,j,k$ subject to a condition
$i+j+k\geq 1$ is understood and
\be
V^2=\frac{(1-w^2(r))^2}{2\beta^2 r^4},\quad K^2=\frac{w'^2(r)}{2\beta^2
r^2},
\ee
and the final factor is the symmetrized trace of the product of even powers
of Pauli matrices:
\be
f(i,j,k) =\Str\left(\tau_1^{2i}\tau_2^{2j}\tau_3^{2k}\right).
\ee
To compute $f(i,j,k)$ explicitly one can derive the following
recurrent relation: \label{eqF}
\bea
(2i+2j+2k)(2i+2j+2k-1)f(i,j,k)&=&2i(2i-1)f(i-1,j,k)+\nonumber\\
  2j(2j-1)f(i,j-1,k)&+&2k(2k-1)f(i,j,k-1).
\eea
When only one index is non-zero one easily finds
\be\label{bcF}
  f(i,0,0)=f(0,i,0)=f(0,0,i)=2.
\ee
The full solution of the Eq.(\ref{eqF}), satisfying the border
conditions
(\ref{bcF}) reads
\begin{equation}\label{Fdef}
  f(i,j,k)=\frac{2(2i)!(2j)!(2k)!(i+j+k)!}{(2i+2j+2k)!i!j!k!}.
\end{equation}
Substituting this explicit expression into the expansion (\ref{Row}) of the
lagrangian one obtains the following representation as a triple series
\be  \label{sum}
L_{NBI}=\frac{\beta^2}{4\pi} \sum_{i+j+k\geq
1}^{\infty}\frac{(-1)^{i+j+k}(2i)!(2j)!(2k)!}
 {i!^2 j!^2 k!^2(2i+2j+2k-1)}
 \left(\frac{V}{2}\right)^{2i}\left(\frac{K}{2}\right)^{2j+2k}.
\ee
Remarkably, one can perform this summation explicitly. First we
observe that, once the factor $1/(2i+2j+2k-1)$ is omitted, and the
summation is extended to all values of power indices including
zero, we obtain a triple series expansion for the function
\be
Z(V,K)=\frac{1}{\sqrt{1+V^2}(1+K^2)}.
\ee
Indeed, treating this function as a product of three square roots,
two of which are equal, we find the following representation:
\be
  Z(V,K)=\sum_{i,j,k=0}^{\infty}\frac{(2i)!(2j)!(2k)!}{i!^2 j!^2 k!^2}
(-1)^{i+j+k}\left(\frac{V}{2}\right)^{2i}\left(\frac{K}{2}\right)^{2j+2k}
 \label{rhrow}
 \ee
Now it is easy to see that the desired sum  (\ref{sum}) is related to
$Z(V,K)$ through the following differential equation
\be
 K\dif{K}L(K,V)+V\dif{V} L(K,V)-L(K,V) =\frac{\beta^2}{4\pi}(Z(V,K)-1), \label{pderow}
\ee
where $L(K,V)$ stands for $L_{NBI}$.  Its solution satisfying the
initial conditions
\be
L(0,0)=0,\quad \dif{V}L (0,0)=\dif{K}L(0)=0,
\ee
following from an initial definition of the lagrangian
(\ref{LNBI}), reads
\be
L_{NBI}=\frac{\beta^2}{4\pi}\left(1-\frac{1+V^2+K^2\cA}{\sqrt{1+V^2}}
\right),
\ee
where
\be\label{cAdef}
\cA=\sqrt{\frac{1+V^2}{V^2-K^2}} \ath\sqrt{\frac{V^2-K^2}{1+V^2}}.
\ee
Here we assumed that $V^2>K^2$, otherwise an $\arctan$ form is
more appropriate. Note that when the difference $V^2-K^2$ changes
sign, the function $\cA$ remains real valued. It can be checked
that when $\beta \to \infty$, the standard Yang-Mills lagrangian
(restricted to monopole ansatz) is recovered. In the strong field
region our expression differs essentially from the square
root/ordinary trace lagrangian.
%%%%%%%%%%%%%%%%%%%%%%%%%%%%%%%%%%%%%%%%%%%%%%%%%%%%%%%%%%%%%%%%%%%%%%
\section{Glueballs}
%%%%%%%%%%%%%%%%%%%%%%%%%%%%%%%%%%%%%%%%%%%%%%%%%%%%%%%%%%%%%%%%%%%%%%
The standard Yang-Mills theory does not admit classical
particle-like solutions \cite{De76,Pa77,Co77}. More precisely,
this famous no-go result asserts that there exist no
finite--energy nonsingular solutions to the four--dimensional
Yang-Mills equations which would be either static, or
non-radiating time--dependent \cite{Co77}. This result follows
from the conformal invariance of the Yang-Mills theory with the
quadratic action, which implies that the stress--energy tensor is
traceless : $T_\mu^\mu=0=-T_{00}+ T_{ii}$, where $\mu=0,...,3,\;
i=1,2,3$. Since $T_{00}>0$, the sum of the principal pressures
$T_{ii}$ is everywhere positive, {\it i.e.} the Yang-Mills matter
is repulsive. This makes the mechanical equilibrium impossible
\cite{Gi82}. In the spontaneously broken gauge theories conformal
invariance is also broken by scalar fields. Thus the above
obstruction is removed what opens the possibility of particle-like
solutions: magnetic monopoles and sphalerons. Monopoles are
essentially related to the presence of the Higgs field with the
$SO(3)$ component, while for sphalerons Higgs can be replaced by
other attractive agent. Recall that the sphaleron was first
obtained in the gauge theory with doublet Higgs~\cite{DaHaNe74}
and its existence was explained by Manton \cite{Ma83} as a
consequence of non--triviality of the {\em third} homotopy group
of the broken phase manifold. Later it was found that similar
solutions arise in the theories without Higgs like
Einstein-Yang-Mills and Yang-Mills with dilaton (for a review
see~\cite{VoGa98}), in all such cases conformal invariance is
broken. Recently it was observed that the Born-Infeld modification
of the Yang-Mills action also breaks conformal
invariance~\cite{GaKe99}, and gives rise to particle-like
solutions. They form a discrete sequence labeled by the number of
nodes of the function $w(r)$, and the lower one-node solution is
similar to the sphaleron of the Weinberg-Salam theory.

In~\cite{GaKe99} the NBI lagrangian was adopted in the
'square-root'--ordinary trace form. Now we are able to perform
similar calculations in the $\Str$ version of the NBI theory. It
is worth noting that to study particle-like solutions in the pure
NBI theory without Higgs an exact in $\alpha'$ form of the
lagrangian is needed since these solutions are formed in the
strong field region. Therefore we assume the following one-dimensional action
\be
S_1=\int \;r^2\left(1-\frac{1+V^2+K^2\cA}{\sqrt{1+V^2}} \right)dr.
\ee
Note that the rescaling $\sqrt{\beta}r\to r $ does not change
the action (an overall factor appearing in the
lagrangian  can be removed by the corresponding rescaling of
time). Therefore without loss of generality $\beta$ can be fixed,
it is convenient to choose $\beta=1/\sqrt{2}$. Then the equation
of motion for $w$ will read
\be \label{eqw}
\frac{d}{dr}\left\{\frac{w'}{2(V^2-K^2)}\left(\frac{K^2\sqrt{1+V^2}}{1+K^2}-
\frac{(2V^2-K^2)\cA}{\sqrt{V^2-K^2}}\right)\right\} =
 \frac{Vw(K^2\cA -V^2)}{(V^2-K^2)\sqrt{1+V^2}}.
\ee
where now
\be
V^2=\frac{(1-w^2(r))^2}{r^4},\quad K^2=\frac{{w'}^2(r)}{ r^2}
\ee
Analyzing the extremal points of $w$ as discussed in \cite{VoGa98}
one finds that $w$ can not have local minima for $0<w<1, \, w<-1$
and can not have local maxima for $-1<w<0,\,w>1$. Thus any
finite energy solution which starts at the origin on the interval $-1<w<1$ lies
entirely within the strip $-1<w<1$. Once $w$ leaves the strip, it
diverges within a finite distance. We are interested in
particle-like solution with finite total energy (mass) given for
the present model by
\be \label{E}
\cE=\int_0^\infty\;r^2\left(\frac{1+V^2+K^2\cA}{\sqrt{1+V^2}}-1
\right)dr.
\ee
Boundary conditions at the origin can be derived combining the
equation of motion with the requirement of convergence of this
integral. Two classes of solutions are possible: one with $w(0)=0$
(leading to embedded $U(1)$ solutions) and $|w(0)|=1$, relevant
for glueballs. Choosing without loss of generality $w(0)=1$, we
find the following series solution near the origin
\be\label{wor}
w=1-br^2+3b^2r^4 \frac{32 b^4+20 b^2 +3}{2(32 b^4+20b^2+15)}+O(r^6).
\ee

To ensure convergence of the total energy (\ref{E}) at infinity, $w$ must
tend to $\pm 1,\,0$. In the case $w(\infty)= 0$  the global solution is
$w\equiv 0$, i.e. an embedded Abelian.
For non-Abelian solutions one has
\be\label{was}
w=\pm \left(1-\frac{c}{r}\right)+O\left(\frac{1}{r^2}\right),
\ee
where $c$ is another free parameter.
The proof of existence of global solutions
starting at the origin as  (\ref{wor}) and approaching
(\ref{was}) at infinity may be given along the lines of
\cite{GaKe99}. The numerical integration shows that the discrete
family of regular solutions exists for which the parameter $b$
takes values shown in Tab.~\ref{tab:param}.

\begin{table}\begin{center}
\begin{tabular}{|l|l|l|}
  \hline $n$ & $\quad b$                     & $\quad M$
\\\hline $1$ & $\quad 1.23736\times 10^2$    & $\quad 1.20240$
\\\hline $2$ & $\quad 5.05665\times 10^3$    & $\quad 1.234583$
\\\hline $3$ & $\quad 1.67739\times 10^5$    & $\quad 1.235979$
\\\hline $4$ & $\quad 7.11885\times 10^6$    & $\quad 1.236046$
\\\hline $5$ & $\quad 4.94499\times 10^8$    & $\quad 1.2360497$
\\\hline $6$ & $\quad 4.52769\times 10^{10}$ & $\quad 1.2360497$
\\ \hline
\end{tabular}
\end{center}
\caption{Values of $b$ and $M$ for first six
solutions}\label{tab:param}
\end{table}

The integer $n$ is equal to the number of zeroes of $w$. The $n=1$
solution is an analog of the sphaleron known in the Weinberg-Salam
theory \cite{DaHaNe74,Ma83}, it is expected to have one decay
mode. Higher odd-$n$ solutions may be interpreted as excited
sphalerons, they are expected to have $n$ decay directions in the
configuration space. Even-$n$ solutions are topologically trivial,
they can be regarded as sphaleronic excitation of the vacuum.
Qualitatively picture is the same as for the 'square root --
ordinary trace form' \cite{GaKe99} but the quantized values of $b$
are rather different in particular, in that case  $b_1=12.7463$.

Numerical solutions are shown on Fig.~\ref{fwstr}, where for
comparison first four solutions with ordinary trace are also
shown. The difference for the main $n=1$ sphaleron in both
theories is rather small, it increases for higher-$n$ solutions
which move closer to the origin (where two models differ
substantially).
%%%%%%%%%%%%%%%%%%%%%%%%%%%%%%%%%%%%%%%%%%%%%%%%%%%%%%%%%%%%%
\section{Gravitating glueballs and black holes}
%%%%%%%%%%%%%%%%%%%%%%%%%%%%%%%%%%%%%%%%%%%%%%%%%%%%%%%%%%%%%
Now let us consider the $\Str$ NBI theory coupled to gravity. The
corresponding action reads: \be \label{eact}
S_{ENBI}=-\frac{1}{16\pi G}\int\left\{R\sqrt{-g}+4G
\beta^2\left(\Str\sqrt{-\det(g_\mn+\beta^{-1}
F_\mn)}-\sqrt{-g}\right) \right\}d^4x, \ee where $R$ is the scalar
curvature and $G$ is the Newton constant. For static spherically
symmetric solutions the metric can be parameterized as follows:
\be \label{met} ds^2=N\sigma^2
dt^2-\frac{dr^2}N-r^2(d\theta^2+\sin^2\theta d\phi).
\ee A
computation of the symmetrized trace for the action (\ref{eact})
is a straightforward generalization of the above procedure for the
flat spacetime, the main difference being in using  curved
metric (\ref{met}) instead of flat. After suitable rescaling, two dimensional
parameters of the theory $G,\,\beta$ combine in one dimensionless
coupling constant
\be
  g=G\beta,
\ee
which is the only substantial parameter of the  theory.
 The reduced one dimensional action reads
\be
S_{1}=\int  \left\{ \frac\sigma
2\left(1+N\left(1+2r\left(\frac{\sigma'}\sigma+
\frac{N'}{2N}\right)\right)\right) +g
r^2\sigma\left(1-\frac{1+K^2+V^2\cA}{\sqrt{1+K^2}}\right)\right\}dr,
\ee
where now
\be
 K^2=\frac{(1-w^2)^2}{r^4},\qquad V^2=\frac{N w'^2}{r^2}, \quad
\ee
and $\cA$ is still defined by (\ref{cAdef}).

The equations of motion derived from this  action consist of
 an equation for the metric function $\sigma$:
\be
\frac{\sigma'}{\sigma}={\frac {g\,{K}^{2}r\left (\left
(2\,{V}^{2}+2\,{K}^{2}{V}^{2}-{K}^{2}-{K}^{4}\right ){
\cA}-{K}^{2}-{K}^{2}{V}^{2}\right )}{N\sqrt {1+{V}^{2}}\left
({V}^{ 2}+{K}^{2}{V}^{2}-{K}^{2}-{K}^{4}\right )}},
\ee
an equation for the local mass function $m(r)$ defined via $N=1-2m/r$:
 \be
  m'=gr^2\left(\frac{1+V^2+\cA\,K^2}{\sqrt
  {1+V^2}}-1\right),
 \ee
and the following equation for $w$:
\be
\frac{d}{dr}\left\{\frac{N\sigma w'}{2(V^2-K^2)}\left( \frac
{{K}^{2}\sqrt {1+{V}^{2}}}{1+{K}^{2}}-\frac
{(2{V}^{2}-{K}^{2})\cA}{\sqrt{1+{V}^{2}}}\right)\right\}=\frac{\sigma
V w(K^2\cA-V^2)}{(V^2-K^2)\sqrt{1+V^2}}.
\ee

The equation for $\sigma$ decouples form the rest of the system, and this
function can be found by a simple integration once $N$ and $w$ are known.
Therefore we concentrate on a coupled system for $N,\,w$ which is obtained
after using the $\sigma$-equation in the $w$-equation.
Like in the ordinary trace version, these equations admit an embedded $U(1)$
solution $\sigma\equiv 1,\; w\equiv 0$ corresponding to the unit magnetic
charge. The corresponding metric was given in \cite{GiRa95,Ra97}
(see also \cite{WiSoKu00}):
\be
N=1-\frac{2}{r}\left(m_0+g\int_0^r\left(\sqrt{1+x^4}-x^2\right)dx\right),
\ee
where $m_0$ is a free parameter which can be positive, zero or negative.
For $m_0=0$ the black hole solutions (with the event horizon) exist for
$g>g_{cr}=1/2$, otherwise there is no horizon.
The role of this critical value in the non-Abelian case
was discussed in \cite{WiSoKu00} and \cite{DyGa00} (for the ordinary trace
$SU(2)$ Born-Infeld action). The metric for regular solutions approaches
that of an Abelian solution (without horizon and $m_0=0$) for $g<g_{cr}$
and large $n$, the limiting mass being the corresponding Abelian mass.

The present model also have regular gravitating solutions for all values of
$g$ which can be thought as interpolating between the flat case
solutions discussed above and Bartnik-McKinnon \cite{BaMc88} solutions of
Einstein-quadratic Yang-Mills model
(for a detailed discussion see \cite{VoGa98}).
The situation is very much alike to the
case of the ordinary trace ENBI model \cite{DyGa00}.

Regular gravitating solutions start at the origin with the following
series expansion
\bea \label{ge}
w&=&1-b\,r^2+O(r^4)\nonumber\\
N&=&1-\frac 23
\frac{g(1+8b^2-\sqrt{1+4b^2})}{\sqrt{1+4b^2}}r^2+O(r^4)
\eea
where $b$ is again a free parameter. At infinity one should have
\be
w\to \pm 1\quad, N\sim 1-\frac{2M}{r},
\ee
where $M$ is the Schwarzschild (ADM) mass. Numerical solutions
interpolating between these asymptotics are shown in
Figs.~\ref{gw},~\ref{gN}. Note that the difference between $\Str$
and $\tr$ solutions is decreased with respect to the flat space
case, especially for the first $n=1$ sphaleron. The metric bending
is slightly more pronounced in the $\Str$ case. In Fig.~\ref{b} we
show the dependence of the parameter $b_1$ for the $n=1$ solution
on the gravitational coupling constant for both $\Str$ and $\tr$
versions of the NBI theory. Qualitatively their behavior is the
same. For $g\to\infty$ both curves converge to the rescaled
Bartnik-McKinnon's value.

In weak gravity limit (as $g\to 0$), $w(r)$-functions for regular solutions
do not differ considerably from the flat case, especially the sequence
$b_n$ is unbounded while the number $n$ of nodes of $w$ tends to infinity.
The metric function $N(t)$ with
increasing $n$ approaches the metric of the abelian solution at some
interval which moves more and more close to the origin. The ADM masses of
these solutions behave like
\be\lim_{g\to 0}\frac{M_n(g)}{g}=M^{flat}_n \ee
where $M^{flat}_n$ are the flat case masses defined by (\ref{E}) and
shown in table 1.

But if $g>g_{cr}=1/2$ the situation changes. With increasing $n$
the parameters $b_n$ tend to a limiting value and the metric
functions tend to the metric of the limiting abelian solution with
a degenerate horizon. This situation resembles that of the
Einstein-Yang-Mills model and indeed this model could be recovered
in the limit $g\to \infty$ after a rescaling $\sqrt g r\to   r$.
In this limit the glueball masses behave like
\be
\lim_{g\to \infty}\frac{M_n(g)}{\sqrt g}= M_n^{BK},
\ee
where $M^{BK}_n$ is the mass of corresponding Bartnick-McKinnon
solution for the Einstein-Yang-Mills model. This behavior could be
observed in Fig~\ref{m}.

Let us discuss the black holes. Instead of (\ref{ge}) now we specify
boundary conditions at the horizon $r=r_h$ using the following series
expansions:
\bea
N&=&N_h'(r-r_h)+O((r-r_h)^2),\nonumber\\
w&=&w_h+\frac{w_h(w_h^2-1)}{\cA r_h^2N'_h}(r-r_h)+O((r-r_h)^2),
\eea
where
\be
N'_h=\frac1{r_h}2gr_h^2(\sqrt{1+K^2}-1)+1.
\ee
Asymptotically flat black hole solutions exist for all horizon
radii $r_h$ and some discrete sequence of $w_h$, labeled by the
number of nodes of $w$ like in the regular case. Qualitative
behavior of $w$  outside the horizon is also the same. Numerical
results for the metric function $N(r)$ of the $n=1,3$ black holes
are shown in Fig.~\ref{gNbh} for a critical $g=1/2$ and different
values of the horizon radius. For non-small $r_h$ the function $N$
is monotonous outside the horizon, while for smaller $r_h$ one
observes a local minimum near the black hole.

%%%%%%%%%%%%%%%%%%%%%%%%%%%%%%%%%%%%%%%%%%%%%%%%%%%%%%%%%
\section{Discussion}
%%%%%%%%%%%%%%%%%%%%%%%%%%%%%%%%%%%%%%%%%%%%%%%%%%%%%%%%%
Two main results of this paper should be noted. First, we obtained
a closed analytical expression for the symmetrized trace version
of the $SU(2)$ NBI action restricted to monopole ansatz. This
expression is particularly simple and can be further used in
various problems. Moreover, similar technique can be applied to
other related field configurations, the results will be given
elsewhere.

Second, we have extended results of \cite{GaKe99} about the existence
of sphaleronic glueballs in the $SU(2)$ NBI theory without Higgs field to the
symmetrized trace version of the NBI model. In both cases the classical scale
invariance of the ordinary Yang-Mills lagrangian is broken by the Born-Infeld
non-linearity, what removes an obstruction for classical glueballs.
We have found that qualitatively the $\Str$ NBI glueball solutions remain
the same, although a  certain difference is
observed in the core. In particular, quantized values of the
parameter determining the derivative of the Yang-Mills field at the origin
are much larger in the $\Str$ case. Nevertheless,
the masses are not very different and with growing node number converge
rapidly to the same limit. This is due to the fact that most of the energy
is localized in the outer core region where the difference between two
models is less pronounced.

When gravity is taken into account, we have demonstrated that there exist
a continuous transition between the flat space glueballs and
Bartnik-McKinnon's particle-like solutions of the Einstein-Yang-Mills
theory with the usual quadratic lagrangian. The latter can therefore be
viewed as the strong gravity limit of  purely gluonic sphalerons
on $D3$ branes. Gravity was shown to further reduce the difference
between the results of the $\Str$ and $\tr$ models. This is well understood,
since for the strong gravity (in units of the Born-Infeld critical field)
both models have the same limit.

We thank A.~Koshelev for a useful discussion.
 This work was supported in part by the RFBR grant 00-02-16306.
%\bibliographystyle{hunsrt}
%\bibliography{bi}

\newpage
\newcommand{\ylabel}[1]{\put(2.2,8.5){\hbox to 0pt{\hss\large #1}}}
\newcommand{\xlabel}[1]{\put(11.2,0.3){\hbox to 0pt{\hss\large #1}}}
%%%%%%%%%%%%%%%%% w for flat glueball Str/Tr
\begin{figure}[p]
\unitlength1cm
\begin{picture}(14,11)
\ylabel{$w(r)$} \xlabel{$\ln r$}
\put(1,0){\epsfig{width=14cm,height=11cm,file=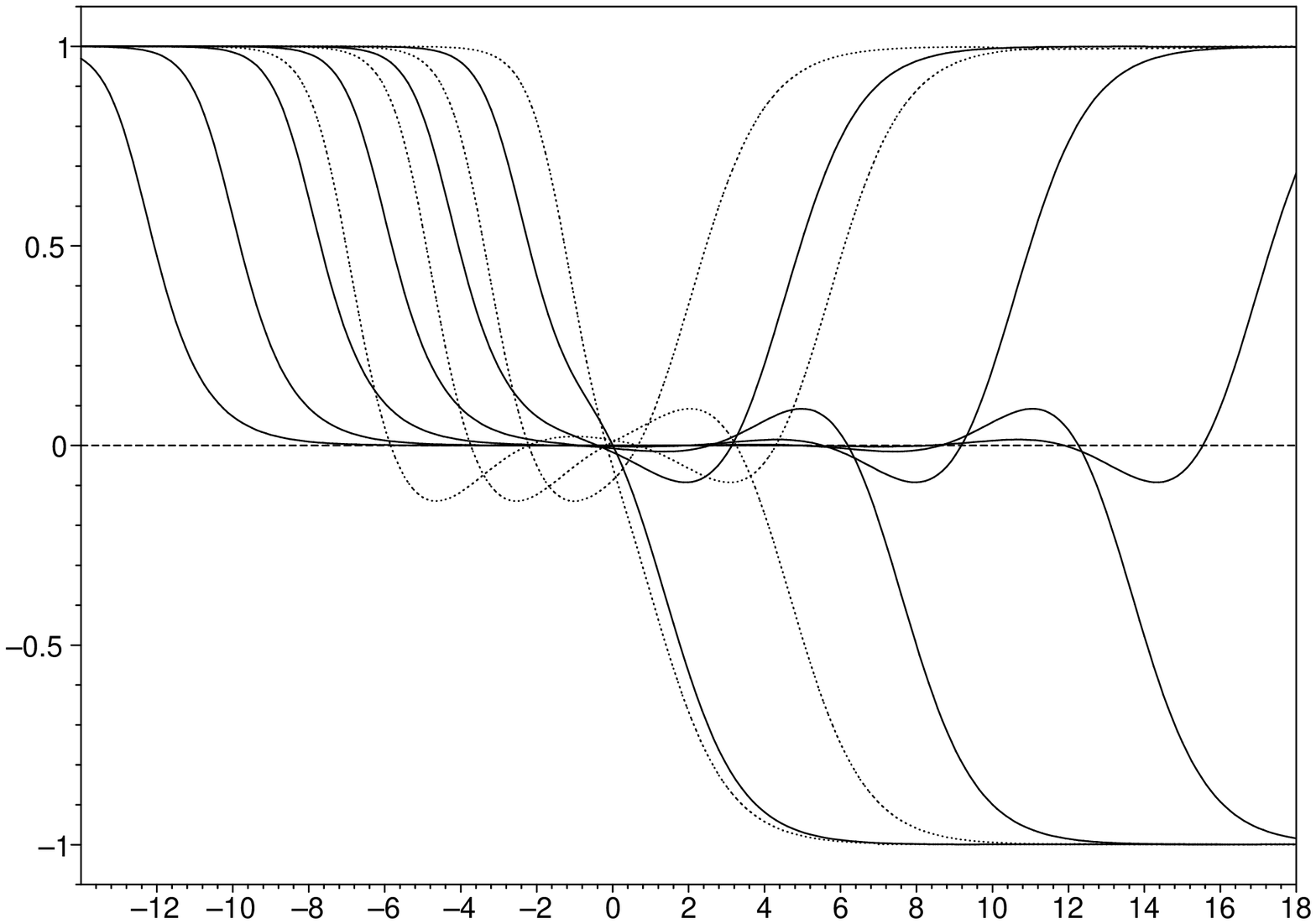}}
\end{picture}
\caption{First six solutions $w_n(r)$ for flat space glueballs in
the NBI theory with symmetrized trace (solid line) and their four
ordinary trace counterparts (dotted line).} \label{fwstr}
\end{figure}
%%%%%%%%%%%%%%%%% w for gravitating glueballs Str/Tr
\begin{figure}[p]
\unitlength1cm
\begin{picture}(14,11)
\ylabel{$w(r)$} \xlabel{$\ln r$} \put(7.7,2.0){\large$n=1$}
\put(9.8,7.0){\large$n=2$} \put(10.8,3.5){\large$n=3$}
\put(1,0){\epsfig{file=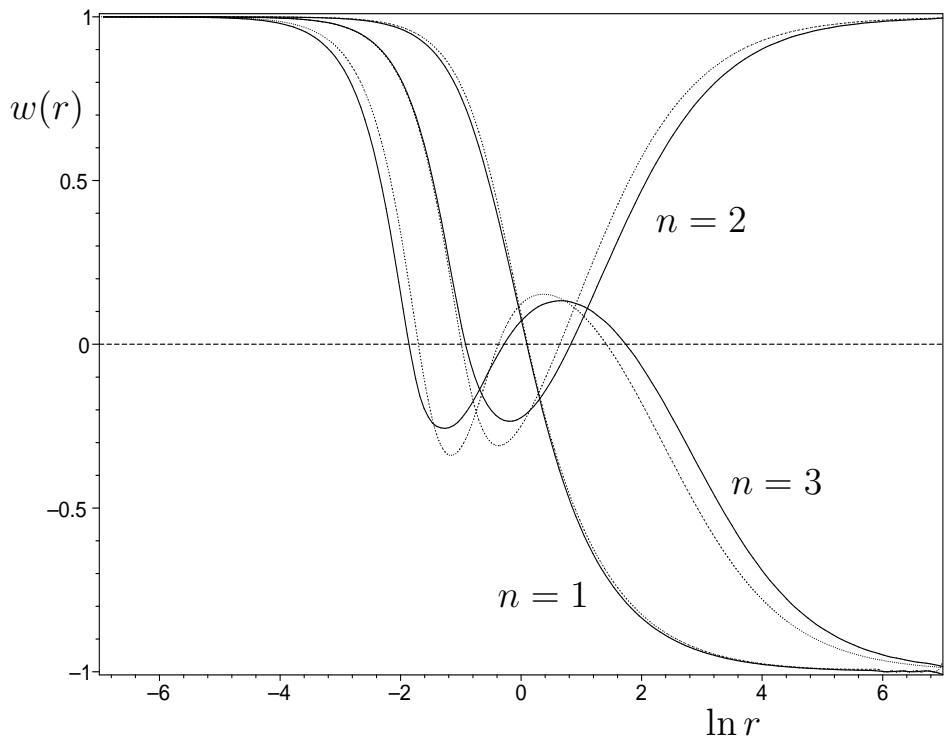,width=14cm,height=11cm}}
\end{picture}
\caption{Functions $w_n(r),\; n=1,2,3$ for gravitating glueballs
in the Str (solid line) and Tr (dotted line) NBI models with
$g=g_{cr}=\frac12$. Solutions are practically indistinguishable
for $n=1$, becoming slightly different for $n=2,3$.} \label{gw}
\end{figure}
%%%%%%%%%%%%%%%%% N for gravitating glueballs Str/Tr
\begin{figure}[p]
\unitlength1cm
\begin{picture}(14,11)
\ylabel{$N(r)$} \xlabel{$\ln r$} \put(7.0,8.3){\large$n=1$}
\put(6.8,4.0){\large$n=2$} \put(4.5,2.5){\large$n=3$}

\put(1,0){\epsfig{file=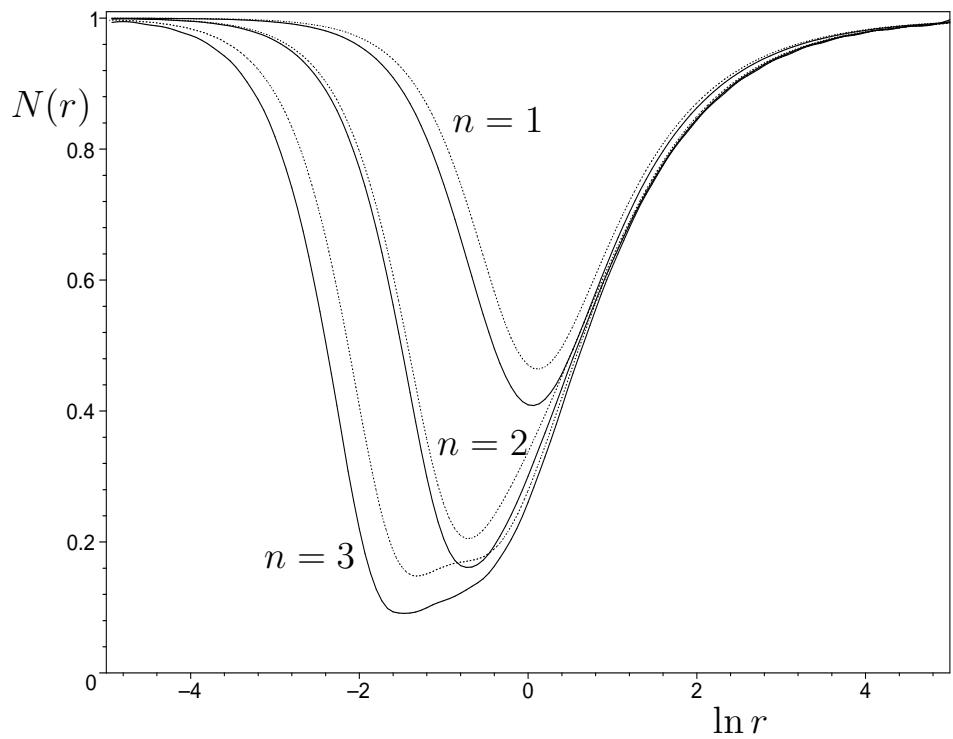,width=14cm,height=11cm}}
\end{picture}
\caption{Metric functions $N_n(r),\; n=1,2,3$ for gravitating
glueballs in the Str (solid line) and Tr (dotted line) NBI models
with $g=1/2$. Gravitational binding in the Str case is slightly
greater than in the Tr case.} \label{gN}
\end{figure}
%%%%%%%%%%%%%%%%% b(g) for n=1 gravitating glueball Str/Tr
\begin{figure}[p]
\unitlength1cm
\begin{picture}(14,11)
\ylabel{$\log_{10} b_1$} \xlabel{$\log_{10} g$} \put(8,8.7){\large
Str} \put(5,6.7){\large tr}
\put(1,0){\epsfig{file=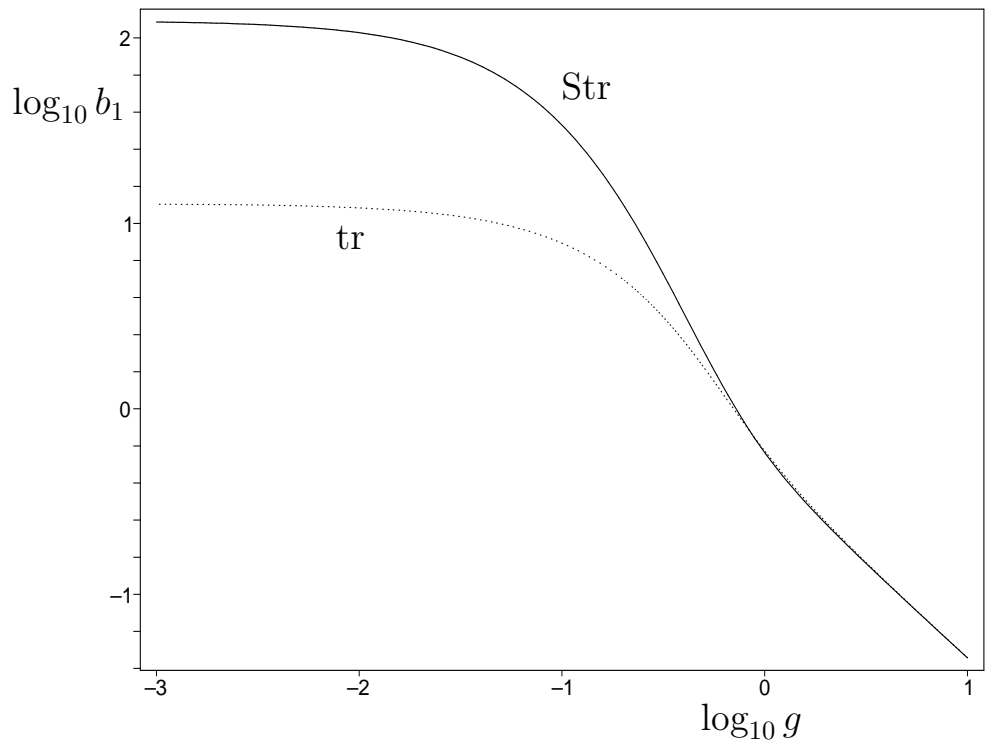,width=14cm,height=11cm}}
\end{picture}
\caption{Parameter $b$ versus an effective coupling $g$ (in $\log$
variables) for the gravitating NBI model $n=1$ solutions in the
Str (solid line) and Tr (dotted line) models.} \label{b}
\end{figure}
%%%%%%%%%%%%%%%%% m(g) for gravitating glueball Str/Tr
\begin{figure}[p]
\unitlength1cm
\begin{picture}(14,11)
\ylabel{$\frac{M_1}{\sqrt{g}}$}
\xlabel{$\log_{10} g$}
\put(8.2,5){\large Str} \put(9,4){\large tr}
\put(13.7,9.55){\large$M_1^{BK}$}
\put(1,0){\epsfig{file=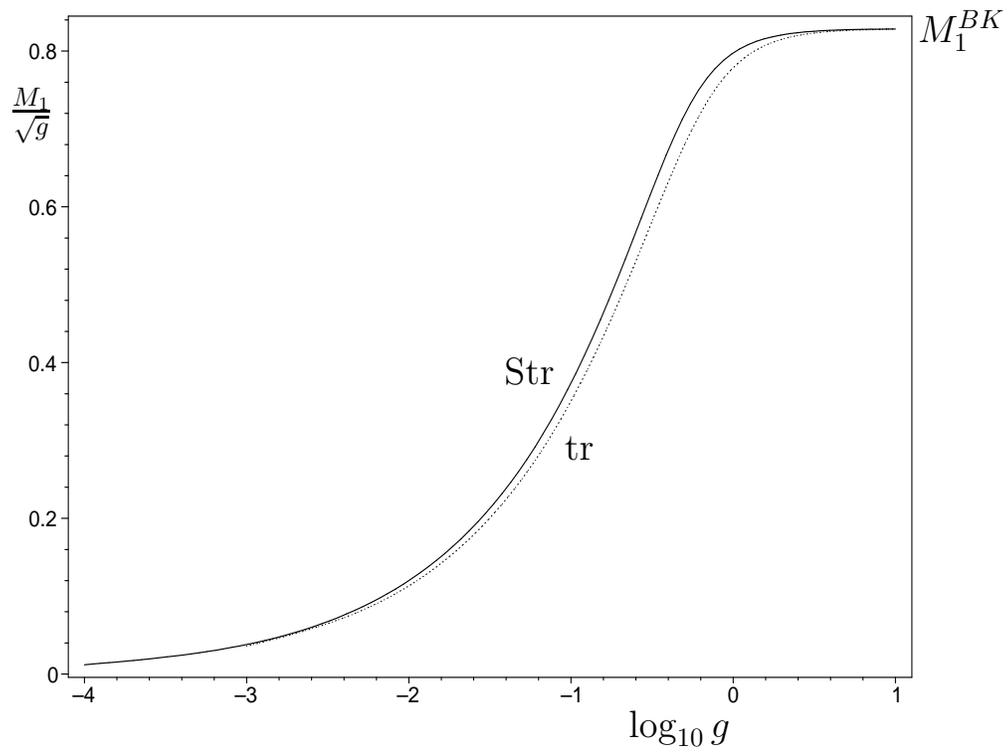,width=14cm,height=11cm}}
\end{picture}
\caption{Dependence of mass on $g$ for the $n=1$ gravitating
glueball (in units $\sqrt{g}$) in the ENBI theory with symmetrized
trace (solid line) and ordinary trace (dotted line).} \label{m}
\end{figure}
%%%%%%%%%%%%%%%%% N for Str black holes
\begin{figure}[p]
\unitlength1cm
\begin{picture}(14,11)
\ylabel{$N(r)$} \xlabel{$\ln r$} \put(3.2,1.7){\large$r_h=0.1$}
\put(7.5,4){\large$r_h=1$} \put(3.5,9.3){\large$r_h=0.01$}

\put(1,0){\epsfig{file=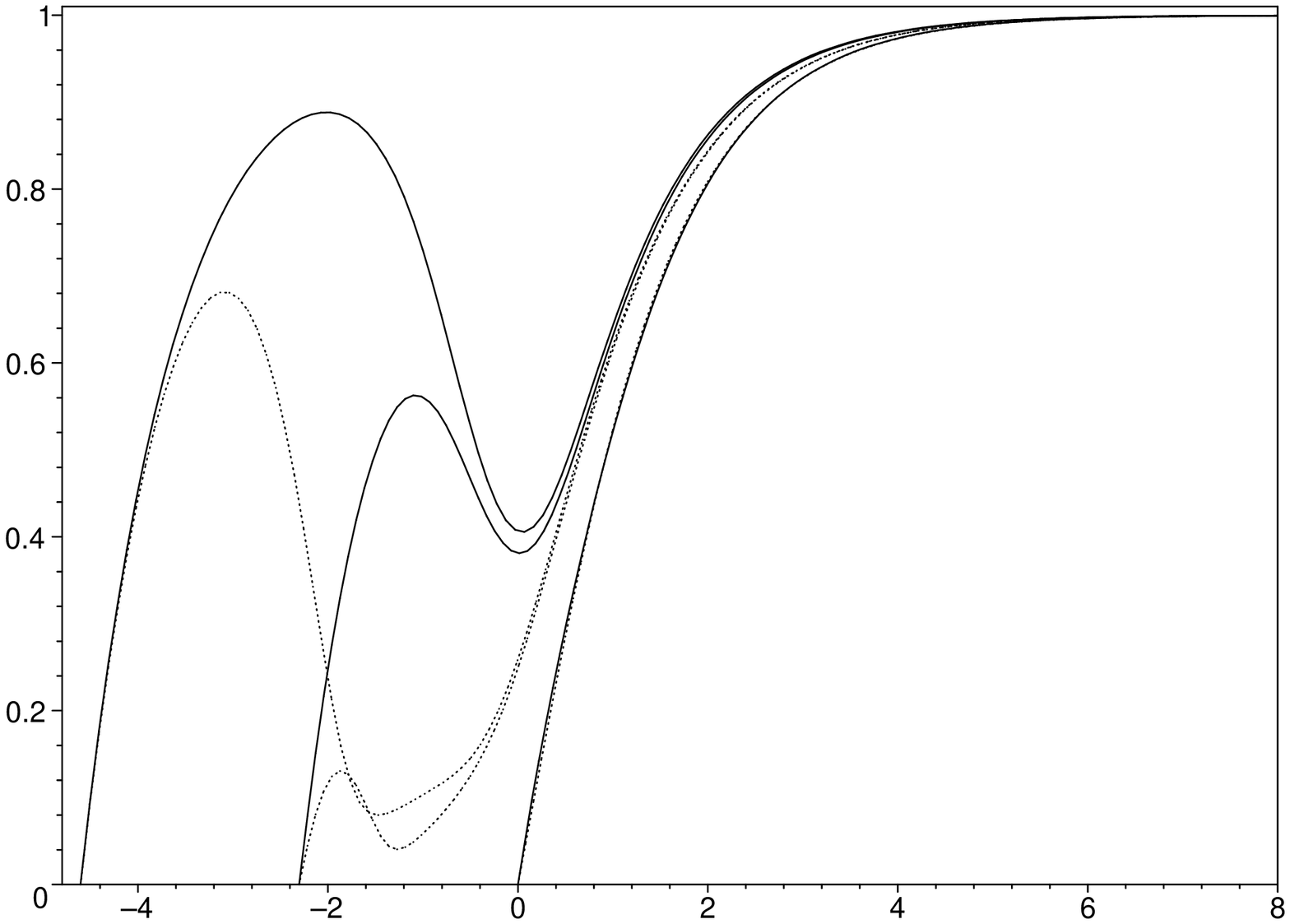,width=14cm,height=11cm}}
\end{picture}
\caption{Metric functions $N_n(r)$ for black holes in the Str
model for $n=1$ (solid line) and $n=3$ (dotted line) with
$r_h=1,0.1,0.01$ and $g=\frac12 $.} \label{gNbh}
\end{figure}

\end{document}